\documentclass{article}
\usepackage{graphicx}
\usepackage{here}
\def\beq{\begin{equation}}
\def\eeq{\end{equation}}
\def\bea{\begin{eqnarray}}
\def\eea{\end{eqnarray}}
\def\bq{\begin{quote}}
\def\eq{\end{quote}}

\def\nnb{\nonumber}
\def\ga{\left(}
\def\dr{\right)}

\def\rar{\rightarrow}

\def\nnb{\nonumber}
\def\la{\langle}
\def\ra{\rangle}
\def\nin{\noindent}
\def\ba{\begin{array}}
\def\ea{\end{array}}

\def\als{\alpha_s}
\def\as{\ga\frac{\bar{\alpha_s}}{\pi}\dr}
\def\asr{\ga\frac{{\alpha_s}}{\pi}\dr}

\begin{document}
\begin{flushright}
PM/01-xx\\ 
\end{flushright}
\vspace*{5mm}
\section*{$c,b$ quark masses and $f_{D_{(s)}},f_{B_{(s)}}$ decay constants from \\
pseudoscalar sum rules in full QCD to order $\alpha_s^2$}
\vspace*{1.cm}
{\bf Stephan Narison}
\\
Laboratoire de Physique Math\'ematique et Th\'{e}orique,
UM2, Place Eug\`ene Bataillon,
34095 Montpellier Cedex 05, France
and
Center for Academic Excellence on Cosmology and Particle Astrophysics (CosPA),
Department of Physics, National Taiwan University, 
Taipei, Taiwan, 10617 Republic of China.\\
Email: narison@lpm.univ-montp2.fr
E-mail:
narison@lpm.univ-montp2.fr\\

\vspace*{.5cm}
\noindent
The pseudoscalar sum rules of the heavy-light quark systems are used for extracting {\it
simultaneously} the $c,b$ quark masses and the decay constants $f_{D_{(s)}},f_{B_{(s)}}$ of the
$D_{(s)},B_{(s)}$ mesons. To order $\alpha_s^2$, one obtains the running quark masses: $\bar
m_c(m_c)=(1.10\pm 0.04)$ GeV,
$\bar m_b(m_b)=(4.05\pm 0.06)$ GeV, the perturbative pole masses:
$M_c=(1.46\pm 0.04)$ GeV, $M_b=(4.69\pm 0.06)$ GeV, and the decay constants: $f_D=(205\pm
20)$ MeV,
$f_B=(203\pm 23)$ MeV and $f_{D_s}=(235\pm 24)~{\rm MeV}~,~f_{B_s}=(236\pm 30)~{\rm MeV}$, in the normalization
where $f_\pi= 130.56$ MeV. The
fitted values of the pole and running masses satisfy quite well their three-loop perturbative
relation. The value $f_D\simeq f_B$ confirms earlier findings from the sum rule that the
$1/\sqrt{M_P}$ heavy quark symmetry scaling law is affected by large $1/M_P$ corrections. 
\noindent
\vspace*{.5cm}
\section{Introduction}
One of the most important parameters of the standard model is the
quark masses. However, contrary to the leptons, where the physical mass
can be identified with the pole of the propagator, the quark masses are
difficult to define because of confinement. Some attempts have been
made in order to define the heavy quark pole mass within perturbation theory,
where it has been shown to be IR-finite \cite{TARRACH} and independent
of the choice of the regularization and renormalization schemes used \cite{SNPOLE}.
More recently, it has been noticed, in the limit of a large number of flavours,
that the resummation of perturbative series can induce a non-perturbative term,
which can affect the truncated perturbative result, 
and can, then, limit the accuracy of
the pole mass determination (for reviews see e.g. \cite{ZAK,BENEKE}). One may bypass the
previous problems, by working, at a given order of
perturbative QCD, with the running quark masses, which are treated like coupling
constants of the QCD Lagrangian (see e.g. \cite{FNR}), and where  
non-perturbative-like effect is expected to be absent. It is
also known that the decay constants $f_{D,B}$ of the pseudoscalar $D,B$ mesons
play an essential r\^ole in the neutral $\bar D$-$D$ and $\bar B$-$B$ oscillations,
and in $D,B$ decays, and can directly be measured in the $D^-, B^-\rar l\nu_l$ leptonic
decays. In addition, it is important for your knowledge of heavy quark symmetry, how the value of
these decay constants deviate from the  na\"\i ve $1/\sqrt{M_P}$ scaling law, expected to occur when the
pseudoscalar meson mass is infinitely large.  A lot of efforts has been furnished in the
literature \cite{PDG} for extracting directly from the data the running masses of the light
 and heavy quarks and the heavy quark ``perturbative" pole masses
using the SVZ QCD spectral sum rules (QSSR) \cite{SVZ} (for a complete review, see e.g.: \cite{SNB}),
whilst
$f_{D,B}$ come from different forms of the pseudoscalar sum rules \cite{SNB}--\cite{JAMIN2} since the
pioneering work of
\cite{NOV}.  In this note, I shall consider a direct extraction of the running charm and bottom quark masses
from pseudoscalar two-point
function sum rules where the $\alpha_s^2$ correction has been recently obtained in \cite{CHET2} and where we shall
also use the observed values of the meson masses: $M_{D^-}=1.869$ GeV and $M_{B^-}=5.279$ GeV. In the same time, 
we shall {\it simultaneously}
estimate the decay constants
$f_{D_{(s)}},f_{B_{(s)}}$ of the $D_{(s)}$ and $B_{(s)}$ mesons. The simultaneous extraction of the quark mass
and
$f_P$ together with the extraction of $f_P$
using the running quark mass
has been initiated in our previous work \cite{SNC,SNFD} to order $\alpha_s$, and will be improved here to order
$\alpha_s^2$. All previous works \cite{SNB}--\cite{SNFD} with the exception of the recent work in
\cite{JAMIN2} have used, as input, the pole mass value for extracting $f_P$, where, as we have mentioned
previously, the definition of the pole mass might be affected by some non-perturbative contributions
\cite{ZAK,BENEKE}. Moreover, the extraction of the quark mass value from the pseudoscalar sum rule
itself which is an improvement of our previous works \cite{SNC,SNFD} is not done in all previous works.
\section{The QCD spectral sum rules}
 We shall work with the pseudoscalar
two-point correlator: 
\beq
\psi_5(q^2) \equiv i \int d^4x ~e^{iqx} \
\la 0\vert {\cal T}
J_q(x)
J^\dagger _q(0) \vert 0 \ra ,
\eeq
built from the heavy-light quark current:
$
J_d(x)=(m_Q+m_d)\bar Q(i\gamma_5)d,
$
and which has the quantum numbers of the $D$ and $B$ mesons. $m_Q$ is the heavy quark mass,
and we shall neglect the $d$ quark mass here.
The corresponding Laplace transform sum rules are:
\beq
{\cal L}(\tau)
= \int_{t_\leq}^{\infty} {dt}~\mbox{e}^{-t\tau}
~\frac{1}{\pi}~\mbox{Im} \psi_5(t),~~~{\mbox {and}}~~~
{\cal R}(\tau) \equiv -\frac{d}{d\tau} \log {{\cal L}(\tau)},
\eeq
where $t_\leq$ is the hadronic threshold. The latter sum  rule,
 or its slight modification, is useful, as it is equal to the 
resonance mass squared, in  
 the simple duality ansatz parametrization of the spectral function:
\beq
\frac{1}{\pi}\mbox{ Im}\psi_5(t)\simeq 2f^2_DM_D^4\delta(t-M^2_D)
 \ + \ 
 ``\mbox{QCD continuum}" \Theta (t-t_c),
\eeq
where the ``QCD continuum comes from the discontinuity of the QCD
diagrams, which is expected to give a good smearing of the
different radial excitations \footnote{At
the optimization scale, its effect is negligible, such that a more
involved parametrization is not necessary.}. The decay constant $f_D$ is
analogous to $f_\pi=92.32$ MeV. However, in order to avoid some confusion, and 
for a more direct comparison
with the lattice results, we shall abandon our favorite normalization, and adopt in the rest of the
paper, the one:
\beq
f_D\equiv\sqrt{2}f_D~,
\eeq
a normalization where $f_\pi=130.56$ MeV, and which the different
experimental groups have also
adopted; 
$t_c$ is the QCD continuum threshold, which is, like the 
sum rule variable $\tau$, an  (a priori) arbitrary 
parameter. In this
paper, we shall impose the
 $t_c$ and $\tau$ stability criteria for extracting our optimal
results \footnote{The corresponding $t_c$ value very roughly indicates
the position of the next radial excitations.}. 
The QCD expression of the correlator
is well-known to two-loop accuracy
(see e.g. \cite{SNB} and the explicit expressions given in \cite{SNFB}),
in terms  of the perturbative pole mass $M_Q$, and including the non-perturbative
condensates of dimensions less than or equal to six
\footnote{A different expression of the coefficient of the quark-gluon mixed
condensate is given in \cite{SOTTO}. This change affects only slightly the result. We shall
include the negligible contribution from the dimension six four-quark condensates. Notice that
there is some discrepancy on the value of the four-quark coefficient in the literature.}. The sum rule
reads:
\bea
{\cal L}(\tau)
&=& M^2_Q\Bigg{\{}\int_{M^2_Q}^{\infty} {dt}~\mbox{e}^{-t\tau}~\frac{1}{8\pi^2}\Bigg{[} 3 t(1-x)^2\ga
1+\frac{4}{3}\asr f(x)\dr+\asr^2 R{2s}\Bigg{]}\nnb\\
&&~\,\, +\Big{[} C_4\la O_4\ra +C_6\la
O_6\ra\tau\Big{]}~\mbox{e}^{-M^2_Q\tau}\Bigg{\}}~,
\eea
where $R{2s}$ is the new $\alpha_s^2$-term obtained semi-analytically in \cite{CHET2} and is available as  a
Mathematica package program Rvs.m. The other terms are:
\bea
x&\equiv& M^2_Q/t,\nnb\\
f(x)&=&\frac{9}{4}+2\rm{Li}_2(x)+\log x \log (1-x)-\frac{3}{2}\log (1/x-1)\nnb\\
& & -\log (1-x)+ x\log (1/x-1)-(x/(1-x))\log x, \nnb\\
C_4\la O_4\ra&=&-M_Q\la \bar dd\ra +\la \als G^2\ra/12\pi\nnb\\
C_6\la O_6\ra&=&\frac{M^3_Q\tau}{2}\ga 1-\frac{1}{2}M^2_Q\tau\dr
g\la\bar d\sigma_{\mu\nu}\frac{\lambda_a}{2}G_a^{\mu\nu}d\ra
\\ &&-\ga\frac{8\pi}{27}\dr\ga 2-\frac{M^2_Q\tau}{2}-\frac{M^4_Q\tau^2}{6}\dr\rho\als \la \bar
\psi\psi\ra^2~.
\eea 
The previous sum rules can be expressed in terms of the running mass $\bar{m}_Q(\nu)$
\footnote{It is clear that, for the non-perturbative terms which are known to leading order
of perturbation theory, one can use either the running or the pole mass. However, we shall see
that this distinction
does not affect notably the present result.},
 through the perturbative  three-loop relation \cite{TARRACH,SNPOLE,GRAY}:
\bea\label{relation}
M_{pole}&=&\bar m(p^2)\Bigg{[}1+\ga\frac{4}{3}+\ln{\frac{p^2}{M^2}}\dr\as+\nnb\\
&&\Bigg{[}K_Q+\ga \frac{221}{24}-\frac{13}{36}n\dr
\ln{\frac{p^2}{M^2}}+\ga\frac{15}{8}-\frac{n}{12}\dr
\ln^2{\frac{p^2}{M^2}}\Bigg{]}\as^2\Bigg{]}~,
\eea ,
where, in the RHS, $M_{pole}\equiv M$ is the pole mass and:
\beq
K_Q=17.1514-1.04137n+\frac{4}{3}\sum_{i\not=Q}\Delta\ga r\equiv 
\frac{m_i}{M_Q}\dr.
\eeq
For $0\leq r\leq 1$, $\Delta(r)$ can be approximated, within an accuracy of 1$\%$
by:
\beq
\Delta(r) \simeq \frac{\pi^2}{8}r-0.597r^2+0.230r^3,
\eeq
Throughout this paper we
shall use the values of the parameters
\cite{SNB,SNG}  given in Table 1. 
\begin{table*}[H]
\begin{center}
\setlength{\tabcolsep}{.28pc}
\caption{Different sources of errors in the estimates of the decay constants (in MeV) and quark
masses (in~GeV). We have exagerately enlarged the error bars of different input in order to have conservative
errors.}
\begin{tabular}{c c c c c c c c c}
\hline 
 & \\
Sources&$|\Delta f_D|_{\bar m_c}$&$|\Delta f_D|_{M_c}$&$|\Delta \bar m_c|$&$|\Delta M_c|$&$|\Delta
f_B|_{\bar m_b}$&$|\Delta f_B|_{M_b}$&$|\Delta \bar m_b|$&$|\Delta M_b|$\\ 
&\\
\hline
&\\
$\Lambda_4=(325\pm  43)$ MeV&7.4&6.2&0.03&0.03 &--&--&--&--\\
$\Lambda_5=(225\pm 30)$ MeV&--&--&--&--&3.6&3.0&0.02&0.02\\
$\nu= M_{c,b}\pm 1/2(M_{c,b}-\bar m_{c,b})$&9.3&--&--&--&14.&--&--&--\\
geom. estimate of $\alpha_s^3$-term&7.9&8.2&--&--&1.7&2.0&--&--\\
$\tau^D=(1.2\pm 0.2)$ GeV$^{-2}$&1.1&1.1&0.01&0.01&--&--&--&--\\
$\tau^B=(0.35\pm 0.05)$ GeV$^{-2}$&--&--&--&--&8.2&5.7&0.02&0.02\\
$6.0\leq t^D_c[\rm{GeV}^2]\leq 9.5$&2.1&2.8&0.01&0.01&--&--&--&--\\
$36.\leq t^B_c[\rm{GeV}^2]\leq 50$&--&--&--&--&2.4&2.8&0.05&0.03\\
$\la \bar dd\ra^{1/3}$(1 GeV)=-$(229\pm 18)$ MeV&8.8&8.9&0.01&0.01&5.2&7.1&--&--\\
$\la \alpha_s G^2\ra=(0.07\pm 0.03)$ GeV$^2$&1.8&1.1&0.005&0.005&0.9&1.&--&--\\
$M^2_0=(0.8\pm 0.1)$ GeV$^2$&0.7&0.7&0.005&0.01&2.2&2.1&0.03&0.04\\
$\alpha_s\la\bar\psi\psi\ra^2=(5.8\pm 2.4)\times 10^{-4}~$GeV$^6$&0.2&0.3&--&--&0.6&1.&0.01&0.01\\
from our estimate of $m_{c,b}$ or $M_{c,b}$&10.9&9.6&--&--&15&13&--&--\\
&\\
Total&20&17 &0.04 &0.04 &23 &17 &0.06&0.06\\
&\\
\hline 
\end{tabular}
\end{center}
\end{table*}
\nin
We have used for the mixed condensate the
parametrization:
\bea
g\la\bar d\sigma_{\mu\nu}\frac{\lambda_a}{2}G_a^{\mu\nu}d\ra&=&M^2_0\la\bar dd\ra,
\eea
and we deduce the value of the QCD scale $\Lambda$ from the value of $\alpha_s(M_Z)=(0.1184\pm 0.031)$
given in \cite{BETHKE,PDG}.
\section{The $D$-meson channel}
We study in Fig. 1, the prediction of $f_D$ from the Laplace sum rules ${\cal L}$
and the one of $M_D$ from ratio of moments
${\cal R}$ for given value of the running charm quark mass $\bar m_c(m_c)$. The influences of the
choice of the continuum threshold $t_c$ and of the sum rule scale $\tau$ are shown in details. Our
optimal results correspond to the case where both stability in $\tau$ and in $t_c$ are reached.
However, for a more conservative estimate of the errors we allow deviations from the stability
points, and we take:
\beq
t_c\simeq (6\sim 9.5)~{\rm GeV}^2~,~~~~~~~~~~~~\tau\simeq (1.2\pm 0.2)~{\rm GeV}^{-2}~,
\eeq
and where the lowest value of $t_c$ corresponds to the beginning of the $\tau$-stability region.
One can inspect
that the dominant non-perturbative contribution  is due to the dimension-four
$M_c\la
\bar dd\ra$ light quark condensate, and test that the OPE is not broken by high-dimension condensates
at the optimization scale. However, the perturbative radiative corrections converge slowly, as the value of
$f_D$ increases by 12\% after the inclusion of the
$\alpha_s$ correction and the sum of the lowest order plus $\alpha_s$-correction increases by 21 \% after the
inclusion of the
$\alpha_s^2$ term, indicating that the total amount of corrections of 21\% is still a reasonnable
correction despite the slow convergence of the perturbative series. However, as the radiative corrections are both
positive, we expect that this slow convergence  will not affect in a sensible way the final estimate. In order to
improve the perturbative contributions, we have estimated the possible contribution of the $\alpha_s^3$-term by
assuming that its coefficient is the geometric sum of the $\alpha_s$ and $\alpha_s^2$ contributions. This effect
is shown in Table 1, which is has a quite reasonnable value. A more precise answer on the higher order
perturbative contribution needs an evaluation of the
$\alpha_s^3$ term which we hope to be available in the near future. In doing the analysis, one can also
notice that the relative size of the perturbative corrections is smaller in the physical observable $f_D$
(perturbative+non-perturbative) than in the perturbative graph alone. 
This is due to fact that the r\^ole of the
$\la\bar\psi\psi\ra$ condensate is important at the optimization scale, which then decreases the
relative weight of the perturbative radiative corrections in the OPE. The estimate of $M_D$ from
the ratio of moments
${\cal R}$ is less affected by radiative corrections, which tend to cancel each others due to the form of
the sum rule. The behaviour of the optimized values of $f_D$ and $M_D$ versus different values of
$\bar m_c(m_c)$ is given in Figs. 
1a and 1b. One can explicitly see in Figs. 1c and 1d that both $M_D$ and
$f_D$ are very sensitive to the change of $\bar m_c(m_c)$. This feature allows to have a good
determination of the quark mass and then of $f_D$, once the experimental value of $M_D$ is used.
Adding quadratically the different errors given in Table 1, we deduce the final estimate:
\beq\label{mcrun}
\bar m_c(m_c)=(1.10\pm 0.04)~{\rm GeV} ~,~~~~~~~~~~~~~~f_D=(201\pm 20)~{\rm MeV}~,
\eeq
where as mentioned previously, we have used the normalization $f_\pi=130.56$ MeV. These optimal
values correspond to $t_c=6.5$ GeV$^2$ and $\tau=1.2$ GeV$^{-2}$. The value of $t_c$ roughly
corresponds to a radial excitation with a mass-splitting relative to the ground state mass of
about $M_\rho$ which is phenomenologically acceptable. 
A similar analysis
shown in Figs. 1e to 1h is done for the pole mass. The discussions presented previously apply also
here, including the one of the radiative corrections. We quote the final result:
\beq\label{mcpole}
 M_c=(1.47\pm 0.04)~{\rm GeV} ~,~~~~~~~~~~~~~~f_D=(208\pm 17)~{\rm MeV}~,
\eeq
where the error is slightly smaller here due to the absence of the subtraction scale uncertainties.
For $f_D$, we consider as a final estimate the mean value of the two predictions and taking the largest
errors:
\beq\label{fdres}
f_D=(204\pm 20)~{\rm MeV}~.
\eeq
Using our previous estimate of $f_{D_s}/f_D=1.15\pm 0.04$ \cite{SNFD}, we can also deduce:
\beq\label{fdsres}
f_{D_s}=(235\pm 24)~{\rm MeV}~.
\eeq
One can immediately compare the present predictions with the one
 obtained to order $\alpha_s$
using the same procedure \cite{SNC}:
\beq
\bar m_c(m_c)=(1.08\pm 0.11)~\mbox{GeV}~,~~~~~~~~~~~~~~f_D=(201\pm 15)~{\rm MeV}~,
\eeq
where one can notice a good agreement between the $\alpha_s$  and $\alpha_s^2$
results, though the error in \cite{SNC} is smaller
as the effect of the subtraction point $\nu$ has not been taken into account. 
However, the agreement seems paradoxal in view of the fact that radiative corrections tend to
increase the value of $f_D$ compared to the lower orders result. The different truncations
of the expression of $\alpha_s$ and of the relation between the pole and running mass also affect the
absolute value of $f_D$, which tend to compensate the increase due to the radiative
corrections of the correlator. Therefore, a na\"\i ve comparison becomes misleading.  As one can see in
Table 1, the main source of errors is due to the variations of the quark mass and to a lesser extent to the
ones of
$\nu$, $\Lambda$ and $\la\bar\psi\psi\ra$. The effect of $t_c$ on the result is relatively
small from the value $t_c$ larger than 6 GeV$^2$, where one starts to have a $\tau $ stability.
This result for the mass is also in agreement with the one from $M_{J/\psi}$ \cite{SNM,PDG}:
\beq
\bar m_c(m_c)=(1.23^{+0.04}_{-0.05})~\mbox{GeV}~,
\eeq
but lower than the one from \cite{PINEDA} using non-relativistic Balmer formula for the $\bar cc$ bound
state. One can cross-check that the two values of $\bar m_c(m_c)$ and $M_c$ give the ratio:
\beq
{M_c}/{\bar m_c(m_c)}\simeq  1.33~,
\eeq
 which satisfies quite well the
three-loop perturbative relation $M_c/\bar m_c(m_c)=1.33$ obtained from the previous Eq.
(\ref{relation}). This could be a non-trivial result if one has in mind that the quark pole mass
definition can be affected by non-perturbative corrections not present in the standard SVZ-OPE. In
particular, it may signal that
$1/q^2$ correction of the type discussed in \cite{ZAK,CNZ}, if present, will only affect weakly the standard
SVZ-phenomenology as observed explicitly in the light quark, gluonia and hybrid channels \cite{CNZ}.
Recent results to order
$\alpha_s^2$
\cite{PENIN}, using the analogous sum rule and using as input the pole mass value, gives $f_D=(195\pm 20)$
MeV. The result is slightly lower than the result given here, though in agreement within the errors. We
understand this slight difference as due to the lower value of the QCD continuum threshold used there, which
corresponds to
$t_c\approx 5.6$ GeV$^2$ if one uses our pole mass value $M_c=1.47$ GeV. As shown in Fig. 1e, this
$t_c$ value is on the boarder of the $\tau$ stability region. It also indicates that the error
introduced by the choice of
$t_c$ could have been underestimated in this result. Fig. 1g also indicates that $f_D$ is quite
sensitive in full QCD to the change of the $M_c$ value contrary to the remark given in \cite{PENIN}.
Quenched and unquenched lattice results for $f_D$ are compiled in \cite{CUICHINI}. The quenched
results range from $(192\pm 18)$ MeV to $(221\pm 17)$ MeV, in fair agreement with our results within
the errors. The two available unquenched results \cite{CPPAC,MILC} lead to the
average:
\beq
f_D^{\rm lat}=(220\pm 20)~{\rm MeV}~,~~~~~~~~~~~~~~~~~f^{\rm lat}_{D_s}=(254\pm 29)~{\rm MeV}~,
\eeq
which is again in agreement within the errors with our previous estimate. Finally, one can also
compare the value of $f_{D_s}$ with the experimental value \cite{OPAL}:
\beq
f_{D_s}^{\rm exp}=(286\pm 60)~{\rm MeV}~,
\eeq
which agrees within 1$\sigma$ with our prediction. Improvements of our predictions for $f_{D_s}$
need an estimate of the ratio $f_{D_s}/f_D$ to order $\alpha_s^2$.
\section{The $B$-meson channel}
We extend the previous analysis to the case of the $B$-meson, which again is an update of our
previous work in
\cite{SNC,SNB,SNFB}. The analysis
is still similar to the one done in the $D$-channel, and is summarized in Fig 2
and in Table 1. Using the running $b$-quark mass, as a free parameter, we obtain
at the optimization scale $\tau=0.375$ GeV$^{-2}$ and $t_c=38$ GeV$^2$:
\beq\label{mbrun}
 \bar m_b(m_b)=(4.05\pm 0.06)~{\rm GeV} ~,~~~~~~~~~~~~~~f_B=(205\pm 23)~{\rm MeV}~,
\eeq
while using the pole mass as a free parameter, we get:
\beq\label{mbpole}
 M_b=(4.69\pm 0.06)~{\rm GeV} ~,~~~~~~~~~~~~~~f_B=(200\pm 17)~{\rm MeV}~,
\eeq
from which we deduce the average:
\beq\label{fbres}
f_B=(203\pm 23)~{\rm MeV}~,
\eeq
where we have taken the largest errors.
One can again cross-check that the two values of $\bar m_b(m_b)$ and $M_b$ lead to 
\beq
{M_b}/{\bar
m_b(m_b)}=1.16~,
\eeq
to be compared with 1.15 from the three-loop perturbative relation in Eq. (\ref{relation}), and
might indirectly indicate the smallness of the $1/q^2$ correction if any. Our result
of $\bar m_b$ can be compared with our previous estimate from $\Upsilon$-systems \cite{SNM}:
\beq
\bar m_b(m_b)=(4.23\pm 0.05)~{\rm GeV}~,
\eeq 
and with similar values from recent estimates \cite{PDG,BMASS}. Our value of the perturbative pole mass is in
agreement within the errors with the one in \cite{PDG,SNM,BMASS} but is again lower than the one in
\cite{PINEDA}. Our value of $f_B$ is in fair agreement with the recent results
$f_B=(206\pm 20)$ MeV obtained in
\cite{PENIN} using HQET sum rules and the one $f_B=(197\pm 23)$
MeV obtained in \cite{JAMIN2} using the Laplace sum rule like in this work. The slight difference
is due to the different appreciations of the continuum threshold
$t_c$ and $\tau$ stability regions in each papers. More specifically, errors related to the choice of $t_c$
at their choice of lower $\tau$-values appear to be underestimated in these works. At such a choice of
low $\tau$-values, the ground state contribution to the sum rule is smaller than in the
present analysis. Unquenched lattice results
\cite{MILC,CPPAC2} give the mean value:
\beq
f_B^{\rm lat}=(198\pm 37)~{\rm MeV}~,~~~~~~~~~~~~~~~~~{f^{\rm lat}_{B_s}}/{f^{\rm lat}_{B}}=1.17\pm
0.03~,
\eeq
where the largest
error has been taken. These values agree with our previous determination in Eq. (\ref{fbres}) and with our
earlier estimate
\cite{SNFD}:
\beq
{f_{B_s}}/{f_{B}}=1.16\pm
0.05~,
\eeq
which has been confirmed from the recent analysis of \cite{JAMIN2}. Combining this $SU(3)$ breaking ratio with
our estimate of $f_B$, one obtains:
\beq\label{fbsres}
f_{B_s}=(236\pm 30)~{\rm MeV}~.
\eeq
The extension of the previous analysis to the
$D^*$ and
$B^*$ channels is in progress.
\section{Summary}
We have updated our previous estimate of the quark masses and decay
constants in
\cite{SNC,SNFB,SNB} using the recent expression \cite{CHET2} of $\alpha_s^2$ corrections for the heavy-light
pseudoscalar correlators. Our results for the  masses in  Eqs. (\ref{mcrun}, \ref{mcpole},
\ref{mbrun}, \ref{mbpole}) and for the decay constants in Eqs. (\ref{fdres}, \ref{fdsres}, \ref{fbres},
\ref{fbsres})
confirm previous estimates to two-loops. The results for the running masses are:
$$
\bar m_c(m_c)=(1.10\pm 0.04)~{\rm GeV}~, ~~~~~~~\bar m_b(m_b)=(4.23\pm 0.05)~{\rm
GeV}~, ~~~~~~~~~~~~\rm{Eqs}~(\ref{mcrun},~\ref{mbrun})~.
$$ 
The pole masses are:
$$
M_c=(1.47\pm 0.04)~{\rm GeV}~,~~~~~~~~~~~~M_b=(4.69\pm 0.06)~{\rm GeV}~,
~~~~~~~~~~~~~~~~~~\rm{Eqs}~(\ref{mcpole},~\ref{mbpole})~.
$$
The decay constants are:
$$
f_D=(204\pm 20)~{\rm MeV}~,~~~~~~~~~~~~~~~f_B=(203\pm 23)~{\rm MeV}~,
~~~~~~~~~~~~~~~~~~~~~\rm{Eqs}~(\ref{fdres},~\ref{fbres})~.
$$
Using our $SU(3)$ breaking prediction on $f_{P_s}/f_P$ \cite{SNFD}, one also deduces:
$$
f_{D_s}=(235\pm 24)~{\rm MeV}~,~~~~~~~~~~~~~f_{B_s}=(236\pm 30)~{\rm MeV}~,
~~~~~~~~~~~~~~~~~~~~~\rm{Eqs}~(\ref{fdsres},~\ref{fbsres})~.
$$
The three-loop corrections tend to push the
values of the decay constants to higher values restoring the slight discrepancy between the sum rules and recent
unquenched lattice values. The resulting equality $f_D\simeq f_B$ confirm earlier findings
from the sum rule \cite{SNFB} indicating large corrections to the
$1/\sqrt{M_P}$ heavy quark symmetry scaling law. Values of the quark masses obtained from the
pseudoscalar sum rules are in agreement with the one from the quarkonia sum rules
\cite{PDG,SNM,BMASS}, but lower than the ones obtained in \cite{PINEDA} from non-relativistic Balmer
formulae. The fitted values of the running and perturbative pole masses satisfy quite well their
three-loop perturbative relation, which may indicate that $1/q^2$-like terms \cite{ZAK,CNZ} have
negligible effect in this channel like in the case of the light quark systems.

\section*{Acknowledgements}
It is a pleasure to thank W-Y. Pauchy Hwang for the hospitality at
CosPA-NTU (Taipei), where part of this work has been done, Abdesslam Arhrib for his help
in running the three-loop package Rvs.m with his newer Mathematica version,  
Matthias Steinhauser on the use of Rvs.m, Arifa Ali Khan for commenting the CP-PAC results
and Valya Zakharov for some communications. 

\begin{figure}[H]
\begin{center}
\includegraphics[width=16cm]{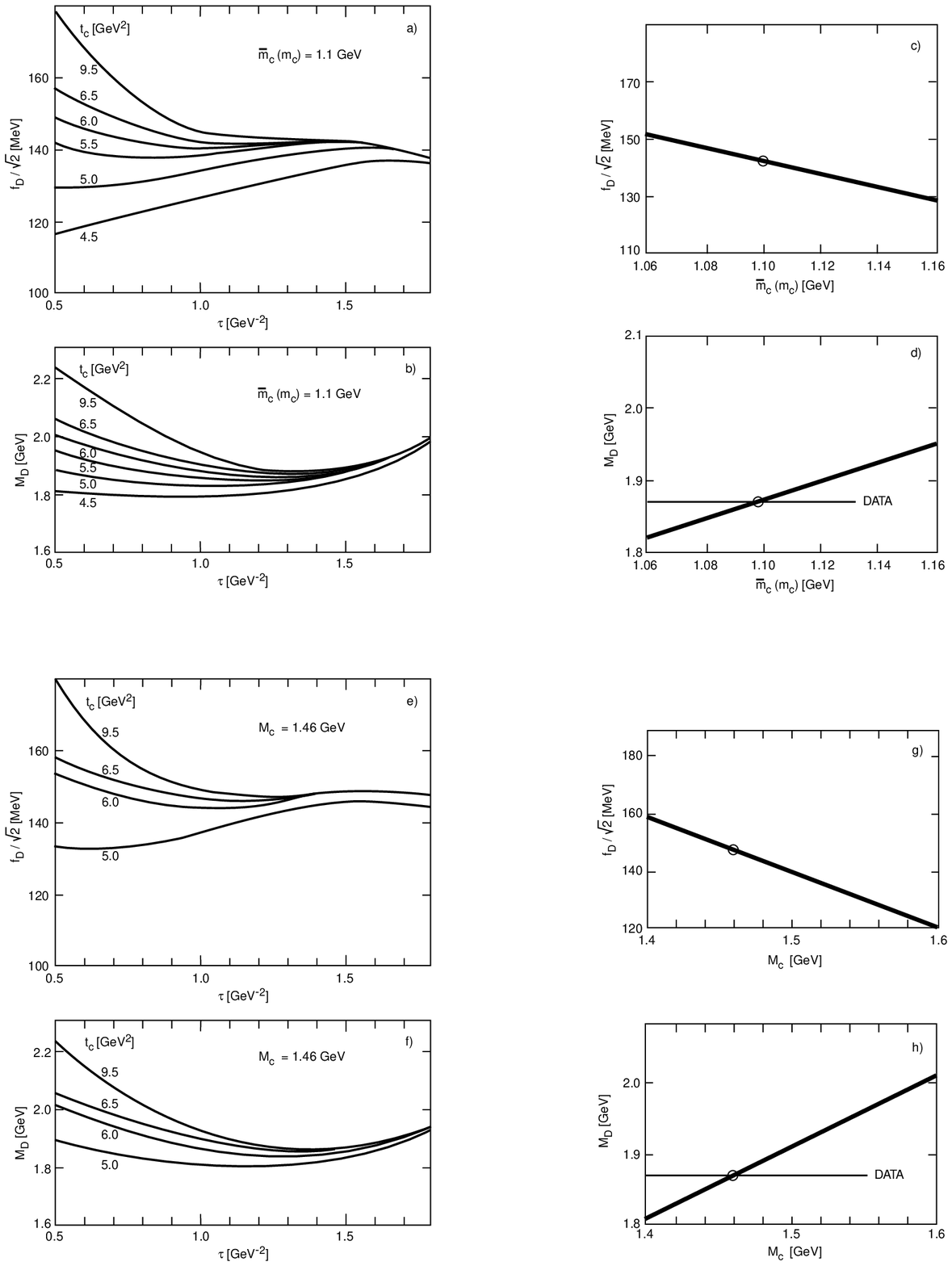}
\caption{Laplace sum rule analysis of the decay constant $f_D$, the running mass $\bar m_c(m_c)$ and
the pole mass $M_c$: a) $f_D$ versus the sum rule scale $\tau$ at given $\bar m_c(m_c)$ and for different
values of the continuum threshold $t_c$; b) the same as a) but for $M_D$; c) and d) effects of $\bar
m_c(m_c)$ on $M_D$ and $f_D$. The circle is the solution given by the data on $M_D$; e) to h) the same as
a) to f) but for the pole mass.}
\end{center}
\end{figure}
\nin
\begin{figure}[H]
\begin{center}
\includegraphics[width=13cm]{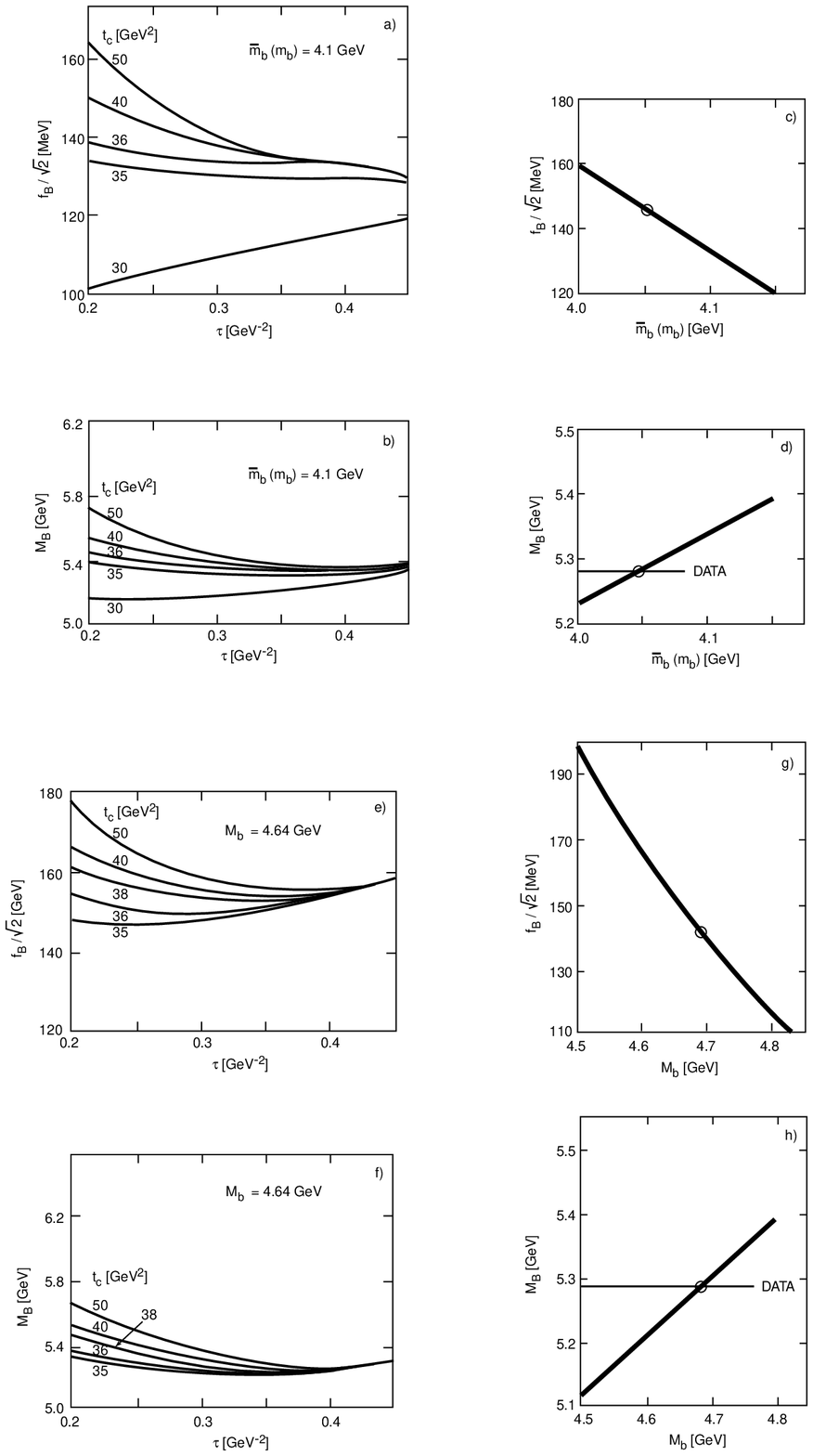}
\caption{The same as Fig. 1 but for the $b$-quark and $B$-meson.} 
\end{center}
\end{figure}
\nin
\end{document}